# A close look at antiferromagnetism in multidimensional phase diagram of electron-doped copper oxide

**Heshan Yu[1], Ge He[1], Ziquan Lin[2], Jie Yuan[1], Beiyi Zhu[1], Yi-feng Yang[1,3], Tao Xiang[1,3], Feo. V. Kusmartsev[4], Liang Li[2], Junfeng Wang[2] and Kui Jin[1,3]**

[1]Beijing National Laboratory for Condensed Matter Physics, Institute of Physics, Chinese Academy of Sciences, Beijing 100190, China

[2]Wuhan National High Magnetic Field Center (WHMFC), Huazhong University of Science and Technology, Wuhan 430074, China

[3]Collaborative Innovation Center of Quantum Matter, Beijing, 100190, China

[4]Department of Physics, Loughborough University, Loughborough LE11 3TU, United Kingdom

E-mail: kuijin@iphy.ac.cn.

**Emergency of superconductivity at the instabilities of antiferromagnetism (AFM), spin/charge density waves has been widely recognized in unconventional superconductors[1-3]. In copper-oxide superconductors, spin fluctuations play a predominant role in electron pairing with electron dopants yet composite orders veil the nature of superconductivity for hole-doped family[4]. However, in electron-doped ones the ending point of AFM is still in controversy for different probes or its sensitivity to oxygen content[5]. Here, by carefully tuning the oxygen content, a systematic study of Hall signal and magnetoresistivity up to 58 Tesla on optimally doped $La_{2-x}Ce_xCuO_{4\pm\delta}$ ($x$ = 0.10) thin films identifies two characteristic temperatures at 62.5±7.5 K and 25±5 K. The former is quite robust whereas the latter becomes flexible with increasing magnetic field, thereby linked to two- and three-dimensional AFM, evident from the multidimensional phase diagram as a function of oxygen as well as Ce dopants[6, 7]. Consequently, the observation of extended AFM phase in contrast to μSR probe[8] corroborates an elevated critical doping in field, providing an unambiguous picture to understand the interactions between AFM and superconductivity.**

In copper oxide superconductors (cuprates), great efforts have been made to attain information on competing orders, on the edge of which the superconductivity usually appears[9, 10]. Multifarious advanced probes as well as transport have been utilized to determine the critical doping levels of correlative Fermi surface reconstruction, quantum critical points, and symmetry broken, etc.[5, 11, 12] However, the precise position of the critical point is still in controversy, obstructing approach to the nature of high-$T_c$ superconductivity. The AFM fades away more slowly in electron-doped cuprates than in hole-doped counterparts, and therefore, determining the boundary of AFM turns out to be a key issue to understand the nature of superconductivity, e.g., the particle-hole symmetry and pairing mechanism.

For electron-doped $Pr_{2-x}Ce_xCuO_4$ and $Nd_{2-x}Ce_xCuO_4$, transport probes present a roughly consistent critical point ($x_{FS}$ ~ 0.16), by the observations of upturn in temperature dependence of resistivity[13], the kink in Hall[14] and Seebeck coefficients[15], the anisotropic in-plane magnetoresistivity[16], the dramatic change in frequency of Shubnikov-de Haas quantum oscillations[17], as well as verified by other probes like angle-resolved photoemission spectroscopy[18, 19] and optical spectra[20]. While, neutron scattering experiment[21] does not observe long range AFM order above $x$ ~ 0.13, before the superconductivity enters.

Very recently, this discrepancy also shows up in $La_{2-x}Ce_xCuO_{4\pm\delta}$ (LCCO), a unique system that has complete phase diagram[6] and optimal doping $x$ ~ 0.10. Previous transport measurements reveal the $x_{FS}$ between 0.13 and 0.15, from the upturn in resistivity[22], Hall coefficient[23] and in-plane angular dependent magnetoresistivity[24]. However, μSR probe only sees long-range AFM till $x$ ~ 0.08, close to the superconducting boundary[8]. The reason for such discrepancy is still obscure, possibly subject to probing sensitivity on different timescale, slightly oxygen variation, field-induced effect, and so on. To figure out this key issue, we performed a systematic electric transport study on optimal Ce-doped $La_{2-x}Ce_xCuO_{4\pm\delta}$ (LCCO, $x$ = 0.1) thin films with fine-tuned oxygen, of which the superconductivity can be suppressed below 10 Tesla. Magnetic field up to 58 Tesla was thus applied to reveal the normal state electronic behavior over a broad region, i.e., resistivity ($\rho_{xx}$), Hall resistivity ($\rho_{xy}$) and coefficient ($R_H$). Consequently, a multidimensional phase diagram ($T$, $B$, $\delta$, $x$) is built up firstly, in conjunction with our previous reported Ce-doping phase. As a result, several key features of AFM in this system are unveiled as follows.

Firstly, $\rho_{xy}$ is not proportional to the magnetic field any more below $T_1$ ~ 62.5 ± 7.5 K. This temperature is quite robust against both the magnetic field ($B$) and oxygen content ($\delta$), in coincidence with the threshold of the anisotropic two-fold in-plane magnetoresistivity (AMR)[24], indicating a mark of intra-plane antiferromagnetism. Secondly, further decreasing the temperature down to $T_2$ ~ 25 ± 5 K, a kink in $R_H(T)$ as well as an upturn of $\rho_{xx}(T)$ is observed after the superconductivity is stripped away. However, this characteristic temperature becomes flexible when $B > 30$ Tesla, suggestive of an inter-plane coupling. Meanwhile, the samples become more conductive in high fields, evident from maxima in $\rho_{xx}(B)$ (at $B_{max}$) and minima in $\rho_{xy}(B)$ (at $B_{min}$). Thirdly, both $B_{max}(T)$ and $B_{min}(T)$ curves display nonmonotonic behavior, and the temperature where the $B_{max}(T)$ or $B_{min}(T)$ reaches extremum displays a positive relation to $T_c(\delta)$, evidencing a link between fluctuations of spin density wave (SDW) and superconductivity. Consequently, the two characteristic temperatures $T_1$ and $T_2$ are naturally associated with two- and three-dimensional AFM, respectively. The fundamental physics in understanding the discrepancy[6-8] on boundary of three-dimension AFM rather lies on a fine controlment of oxygen content, as well as influence from the magnetic field.

Our LCCO thin films were grown on (001)-oriented $SrTiO_3$ substrates by pulsed laser deposition (PLD)[24]. To achieve slightly variation of oxygen content in samples, we carefully controlled the annealing process. The lattice structure of all samples with different annealing time was carefully checked by x-ray diffractometer. Our φ scans

(Fig. 1a-c) and reciprocal space mappings (Fig. 1d-f) demonstrate high quality epitaxial growth, and no impurity phase has been observed from θ/2θ scans (Supplementary Fig. 1). In zero field, the optimized (OP1) sample shows the highest $T_c$, while both the under-annealed (UD1) and over-annealed (OD1) samples exhibit slightly lower $T_c$ (Fig. 1g-i). The UD1 shows a tiny upturn in $\rho_{xx}(T)$ plot above $T_c$, comparable to the underdoped LCCO with $x < 0.10$. Although the metallic behavior remains down to $T_c$, an upturn still appears in 15 Tesla, right at (below) the zero-field superconducting transition for OP1 (OD1).

It is difficult to determine tiny oxygen variation in thin films, while, lattice parameter (*c*-axis) can reflect the change of oxygen content qualitatively[25]. In Fig. 2a, the full superconducting transition temperature $T_{c0}$ is plotted as a function of *c* axis. From the under- to over-annealed samples, the *c* axis gradually shrinks, and the oxygen dependence of superconducting transition, $T_{c0}(\delta)$, exhibits a similar dome-like behavior as the Ce doping $T_{c0}(x)$. As mentioned above, $\rho_{xy}$ is proportional to $B$ at $T > T_1$, a feature of either simplified single band or compensated two bands in metals[26, 27]. Below $T_1 \sim 62.5 \pm 7.5$ K, $\rho_{xy}(B)$ is not linear any more as seen in Fig. 2b. The characteristic temperature $T_1$ is independent on field up to 58 Tesla, and overlaps with each other for all the samples within the error bars (Fig. 2c and Supplementary Fig. 2), implying a robust in-plane AFM coupling of hundred meV (ref. 8 and 24). At lower temperatures, $\rho_{xy}(B)$ changes more dramatically, manifested as a kink in $R_H(T)$ at $T_2 \sim 25 \pm 5$ K (Fig. 2d), roughly following the temperature where the upturn of resistivity appears, reflecting loss of conducting carriers[22]. The $T_2(B, \delta)$ is not so robust as the $T_1(B, \delta)$, mainly attributed to a much weaker inter-plane exchange coupling ($J_\perp$) of ~1-2 meV. Therefore, a self-consistent multidimensional phase diagram is well established, comprised of both $(T, x)$ and $(T, \delta)$ panels (Fig. 3). It is clear that $T_1$ and $T_2$ in $(T, \delta, x = 0.10)$ intersect with the $(T, \delta = 0, x)$ at the starting points of two-fold in-plane AMR and resistivity upturn, corresponding to 2D and 3D AFM, respectively.

All the transport probes[6, 23, 24] point to a static AFM order beyond $x = 0.10$, while the zero-field low energy μSR probe only detects static magnetism below $x = 0.08$ (ref. 8). Our results do show a change of $T_2$ for samples with different oxygen content (Fig. 2c), but such small variation cannot account for the big difference. Since the transport measurements were always done under magnetic field, influence by field should be considered to reconcile this discrepancy. It has been predicted that the competition between spin density wave (SDW) order and superconductivity shifts the quantum critical point to lower doping level, and this point moves back when the superconductivity is destroyed by the magnetic field[28]. Such a plausible explanation implies an elevated AFM critical doping as the superconductivity is being suppressed by magnetic field. Intriguingly, the two quantum critical points, at the boundaries of antiferromagnetism[8] and Fermi liquid[7], move towards opposite directions in fields as illustrated in the inset of Fig. 3, inspiring theoretical considerations in future.

It is obvious that $\rho_{xy}$ shows a minimum with increasing the magnetic field at low temperatures (Fig. 2b and 4a). Meanwhile, there is a maximum in $\rho_{xx}(B)$ curve (Fig. 4b). The $B_{min}(T)$ and $B_{max}(T)$, where minimum of $\rho_{xy}(B)$ and maximum of $\rho_{xx}(B)$ appear, first gradually drop and then quickly increase with raising the temperature (Fig. 4c and 4d, Supplementary Fig. 5 ). Such behavior implies that a SDW gap[29] is closed and then fluctuations dominate the transport, mimicking the manner of closing an energy gap, e.g., suppression of superconducting energy gap followed by amplitude fluctuations in Nernst experiments for both electron- and hole-doped cuprates[30, 31]. An important finding is that the minima of $B_{min}(T)$ and $B_{max}(T)$ show a positive relation to $T_{c0}$ (Fig. 4e and 4f), suggestive of a close relation between SDW fluctuations and superconductivity. Previous work on LCCO has veiled link between spin fluctuations and superconductivity, evident from a positive relation between the linear resistivity and $T_{c0}$. Our results thus provide a route to probe the relation between SDW and superconductivity more directly.

The analysis of the data provides strong evidence of the spin-fluctuations enhancement arising with doping. Owing to strong correlations or large Hubbard $U$ there in $CuO_2$ plane with each doped electron, a magnetic spin-polaron is arising. The spin polaron is located on more than one Cu site having the same spin orientation and surrounded by localized spins of opposite polarity. Therewith the AF structure is not significantly destructed and can be maintained and preserves to larger values of doping. Their appearance can strongly enhance the spin fluctuations and provide the pairing attraction between doped electrons. The spin polarons are different, although play the same role as hole spin-bags proposed long ago for hole-doped cuprates[32], where the relation between the superconducting, $\Delta_{SC}$ and the SDW gap, $\Delta_{SDW}$, has been given: $\Delta_{SC} = \Delta_{SDW} \exp[-\frac{1}{N(E_{k_F})U}]$, here $N(E_{k_F})$ is the density of states on the Fermi energy and $U$ is the pairing attraction, which depends on the binding energy of the spin-polarons and their attraction to each other. Besides the enhancement of spin fluctuations they may also lead to the frustration in the AF interaction between spins and the destruction of the AF long-range order. With increasing temperature, three-dimensional AF order vanishes and the spin-liquid like state forms. In the spin liquid the AF order can be only seen on the short range scale as has been noticed in the present experiments. We observe that such a transition from 3D AFM into the spin liquid state happens when the temperature exceeds $T > T_2$. This critical temperature $T_2$ shifts slightly with tuning the oxygen content in our samples indicating the importance of the electron-spin-bags in the transition from 2D to 3D AF state. Above this temperature we expect to have only the order in the plane which may have a spin liquid character. Finally when the temperature exceeds the higher temperature $T > T_1$ the completely paramagnetic state is formed. Note that $T_1$ does not depend on tuning the oxygen (see Fig. 3) indicating on the importance of the fundamental exchange spin-spin interaction for this transition.

The results of our analysis are presented in the form of holographic phase diagram which has generic features for all electron doped cuprates. This holographic diagram is given in the temperature, magnetic field, Ce, and oxygen doping space, $(T, B, x, \delta)$, respectively. In our analysis we used the data obtained from various underdoped, optimally doped and over-doped samples having different both oxygen and Ce doping. The existence of the spin-bags in the electron doped cuprates indicates on the crucial role of the spin fluctuations in the mechanism of the superconductivity in these strongly correlated systems and naturally explains the connection between the superconducting critical temperature, superconducting gap and the spin density wave gap for the whole phase diagram.

## Reference

1. Scalapino, D.J. A common thread: The pairing interaction for unconventional superconductors. *Rev. Mod. Phys.* **84**, 1383-1417 (2012).
2. Löhneysen, H.v., Rosch, A., Vojta, M. & Wölfle, P. Fermi-liquid instabilities at magnetic quantum phase transitions. *Rev. Mod. Phys.* **79**, 1015-1075 (2007).
3. Chang, J. *et al.* Direct observation of competition between superconductivity and charge density wave order in $YBa_2Cu_3O_{6.67}$. *Nature Phys.* **8**, 871-876 (2012).
4. Taillefer, L. Scattering and Pairing in Cuprate Superconductors. *Annu. Rev. Condens. Matter Phys.* **1**, 51-70 (2010).
5. Armitage, N.P., Fournier, P. & Greene, R.L. Progress and perspectives on electron-doped cuprates. *Rev. Mod. Phys.* **82**, 2421-2487 (2010).
6. Jin, K., Butch, N.P., Kirshenbaum, K., Paglione, J. & Greene, R.L. Link between spin fluctuations and electron pairing in copper oxide superconductors. *Nature* **476**, 73-75 (2011).
7. Butch, N.P., Jin, K., Kirshenbaum, K., Greene, R.L. & Paglione, J. Quantum critical scaling at the edge of Fermi liquid stability in a cuprate superconductor. *Proc. Natl. Acad. Sci. USA* **109**, 8440–8444 (2012).
8. Saadaoui, H. *et al.* The phase diagram of electron-doped $La_{2-x}Ce_xCuO_{4-\delta}$. *Nature Commun.* **6**, 6041 (2015).
9. Keimer, B., Kivelson, S.A., Norman, M.R., Uchida, S. & Zaanen, J. From quantum matter to high-temperature superconductivity in copper oxides. *Nature* **518**, 179-186 (2015).
10. Sachdev, S. Quantum Criticality: Competing Ground States in Low Dimensions. *Science* **288**, 475-480 (2000).
11. Damascelli, A., Hussain, Z. & Shen, Z.-X. Angle-resolved photoemission studies of the cuprate superconductors. *Rev. Mod. Phys.* **75**, 473 (2003).
12. Xia, J. *et al.* Polar Kerr-Effect Measurements of the High-Temperature $YBa_2Cu_3O_{6+x}$ Superconductor: Evidence for Broken Symmetry near the Pseudogap Temperature. *Phys. Rev. Lett.* **100**, 127002 (2008).
13. Fournier, P. *et al.* Insulator-Metal Crossover near Optimal Doping in $Pr_{2-2x}Ce_xCuO_4$: Anomalous Normal-State Low Temperature Resistivity. *Phys. Rev. Lett.* **81**, 4720 (1998).
14. Dagan, Y., Qazilbash, M., Hill, C., Kulkarni, V. & Greene, R. Evidence for a Quantum Phase Transition in $Pr_{2-x}Ce_xCuO_{4-\delta}$ from Transport Measurements. *Phys. Rev. Lett.* **92**, 167001 (2004).
15. Li, P., Behnia, K. & Greene, R. Evidence for a quantum phase transition in electron-doped $Pr_{2-x}Ce_xCuO_{4-\delta}$ from thermopower measurements. *Phys. Rev. B* **75**, 020506(R) (2007).
16. Yu, W., Higgins, J., Bach, P. & Greene, R. Transport evidence of a magnetic quantum phase transition in electron-doped high-temperature superconductors. *Phys. Rev. B* **76**, 020503(R) (2007).
17. Helm, T. *et al.* Evolution of the Fermi Surface of the Electron-Doped High-Temperature Superconductor $Nd_{2-x}Ce_xCuO_4$ Revealed by Shubnikov–de Haas Oscillations. *Phys. Rev. Lett.* **103**, 157002 (2009).
18. Armitage, N.P. *et al.* Doping Dependence of an n-Type Cuprate Superconductor Investigated by Angle-Resolved Photoemission Spectroscopy. *Phys. Rev. Lett.* **88**, 257001 (2002).
19. Matsui, H. *et al.* Evolution of the pseudogap across the magnet-superconductor phase boundary of $Nd_{2-x}Ce_xCuO_4$. *Phys. Rev. B* **75**, 224514 (2007).
20. Onose, Y., Taguchi, Y., Ishizaka, K. & Tokura, Y. Charge dynamics in underdoped $Nd_{2-x}Ce_xCuO_4$:

Pseudogap and related phenomena. *Phys. Rev. B* **69**, 024504 (2004).

21. Motoyama, E.M. *et al.* Spin correlations in the electron-doped high-transition-temperature superconductor $Nd_{2-x}Ce_xCuO_{4\pm\delta}$. *Nature* **445**, 186-189 (2007).
22. Jin, K. *et al.* Normal-state transport in electron-doped $La_{2-x}Ce_xCuO_4$ thin films in magnetic fields up to 40Tesla. *Phys. Rev. B* **77**, 172503 (2008).
23. Jin, K., Zhu, B., Wu, B., Gao, L. & Zhao, B. Low-temperature Hall effect in electron-doped superconducting $La_{2-x}Ce_xCuO_4$ thin films. *Phys. Rev. B* **78**, 174521 (2008).
24. Jin, K., Zhang, X., Bach, P. & Greene, R. Evidence for antiferromagnetic order in $La_{2-x}Ce_xCuO_4$ from angular magnetoresistance measurements. *Phys. Rev. B* **80**, 012501 (2009).
25. Jin, K. *et al.* Anomalous enhancement of the superconducting transition temperature of electron-doped $La_{2-x}Ce_xCuO_4$ and $Pr_{2-x}Ce_xCuO_4$ cuprate heterostructures. *Phys. Rev. B* **83**, 060511(R) (2011).
26. Dong, X. *et al.* $(Li_{0.84}Fe_{0.16})OHFe_{0.98}Se$ superconductor: Ion-exchange synthesis of large single-crystal and highly two-dimensional electron properties. *Phys. Rev. B* **92**, 064515 (2015).
27. Xiang, T., Luo, H.G., Lu, D.H., Shen, K.M. & Shen, Z.X. Intrinsic electron and hole bands in electron-doped cuprate superconductors. *Phys. Rev. B* **79**, 014524 (2009).
28. Sachdev, S. Where is the quantum critical point in the cuprate superconductors? *Phys. Status Solidi B* **247**, 537-543 (2010).
29. Li, P., Balakirev, F. & Greene, R. High-Field Hall Resistivity and Magnetoresistance of Electron-Doped $Pr_{2-x}Ce_xCuO_{4-\delta}$. *Phys. Rev. Lett.* **99**, 047003 (2007).
30. Tafti, F.F. *et al.* Nernst effect in the electron-doped cuprate superconductor $Pr_{2-x}Ce_xCuO_4$: Superconducting fluctuations, upper critical field $H_{c2}$, and the origin of the $T_c$ dome. *Phys. Rev. B* **90**, 024519 (2014).
31. Chang, J. *et al.* Decrease of upper critical field with underdoping in cuprate superconductors. *Nature Phys.* **8**, 751-756 (2012).
32. Schrieffer, J.R., Wen, X.G. & Zhang, S.C. Spin-bag mechanism of high-temperature superconductivity. *Phys. Rev. Lett.* **60**, 944-947 (1988).

**Acknowledgements**

We would like to thank R.L. Greene, J. Paglione, L. Shan, S.L. Li, Z.Y. Meng for fruitful discussions. H.Y thanks L.H. Yang for assistance in structural characterizations. This research was supported by the National Key Basic Research Program of China (2015CB921000), the National Natural Science Foundation of China Grant (11474338), and the Strategic Priority Research Program (B) of the Chinese Academy of Sciences (XDB07020100).

**Figure 1 | Structure characterizations and resistivity for LCCO ($x$=0.1) thin films. a-c,** The φ scans of (103) plane for samples UD1, OP1 and OD1. There are four peaks of nearly equal height, reflecting the high quality of our samples. **d-f,** The reciprocal space mapping of (103) plane of $SrTiO_3$ (cross red pattern) as well as (109) plane of LCCO (oval red pattern). **g-l,** Temperature dependence of resistivity at 0 T and 15 T with $B//c$. An upturn in $\rho(T)$ is observed at a characteristic temperature $T_2$ after the superconductivity is stripped away.

**Figure 2 | Lattice parameter and Hall signal in field. a,** Relation between $T_{c0}$ and $c$-axis lattice parameter. The shape of the $T_{c0}(c)$ behaves as a dome, similar to the Ce-doped superconducting phase. **b,** Magnetic field dependence of Hall resistivity $\rho_{xy}$ from 1.8 K to 85 K. The Hall resistivity is proportional to field at 85 K, but starts to deviate from the linearity at a lower temperature $T_1$. **c,** Two characteristic temperatures in electron-doped $La_{2-x}Ce_xCuO_{4\pm\delta}$ thin films: $T_1$ (triangle symbols) at 62.5 ± 7.5 K is quite robust with increasing magnetic field for all the samples; another characteristic temperature $T_2$ at 25 ± 5K, marking the upturn in resistivity, becomes flexible above 30 T. **d,** Temperature dependence of Hall coefficient $R_H$ shows a kink between 20 K and 30 K, roughly consistent with $T_2$.

**Figure 3 | Multidimensional phase diagram of LCCO as a function of Ce and oxygen.** Along Ce doping axis, the AFM regime (blue area) is achieved by transport measurement in magnetic field. Green spheres denote the boundary of two-dimensional AFM regime, by the in-plane angular magnetoresistance measurements[6]. Blue triangles represent $T_2$. Along the $c$ axis (oxygen doping), for the optimal doping LCCO ($x$=0.1) thin films, the two-dimensional AFM boundary ($T_1$, red spheres) is quite robust against oxygen content. However, the boundary of the three-dimensional AFM, manifested as a kink in $R_H(T)$ curves (red triangle symbols), shifts slightly with tuning the oxygen content. The inset is a sketch of Ce-dependent phase diagram in zero field. There are two quantum critical points (QCPs) at the boundaries of antiferromagnetism[8] and Fermi liquid[7], respectively, which move in opposite directions as indicated by arrows.

**Figure 4 | Correlation between spin density wave and superconductivity. a,** $\rho_{xy}(B)$ from 4.2 K to 26 K shows a minimum with increasing field, where the magnetic field is defined as $B_{min}$. The dash line is a guide to eye. **b,** $\rho_{xx}(B)$ exhibits positive and then negative MR with increasing field, from 4.2 K to 26 K. The red dash line separates the positive- and negative-MR regimes. The magnetic field, corresponding to maximum in $\rho_{xx}(B)$ curves is marked as $B_{max}$. **c,** Temperature dependence of $B_{min}$ for two optimized samples (OP1 and OP2). As lifting up the temperature, $B_{min}(T)$ first gradually decreases and then increases rapidly, suggestive of a closed spin-density-wave gap and enhanced

fluctuations. The temperature where $B_{min}(T)$ shows a minimum is marked as $T_{min}$. **d,** Temperature dependence of $B_{max}$ for OP1 and OP2. The temperature where $B_{max}(T)$ shows a minimum is marked as $T_{max}$. **e** and **f,** $T_{min}$ and $T_{max}$ displays a positive relation to the superconducting transition temperature for five samples UD1, OP1, OP2, OD1 and OD2.

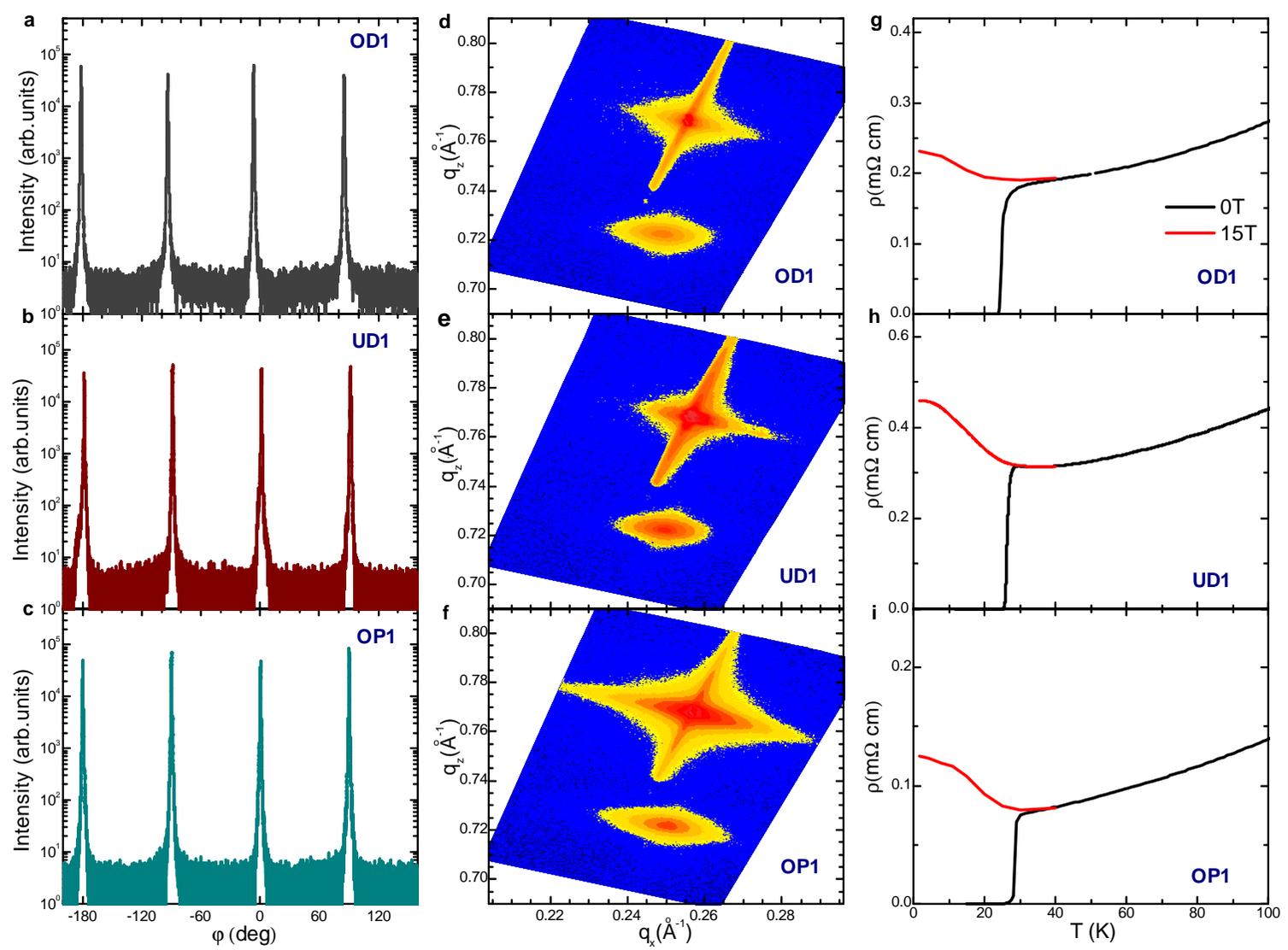

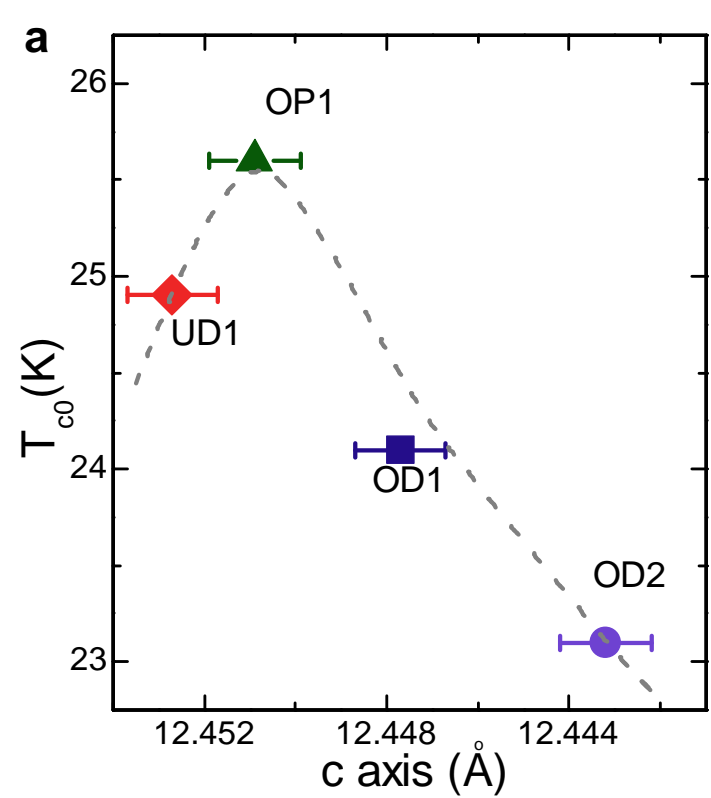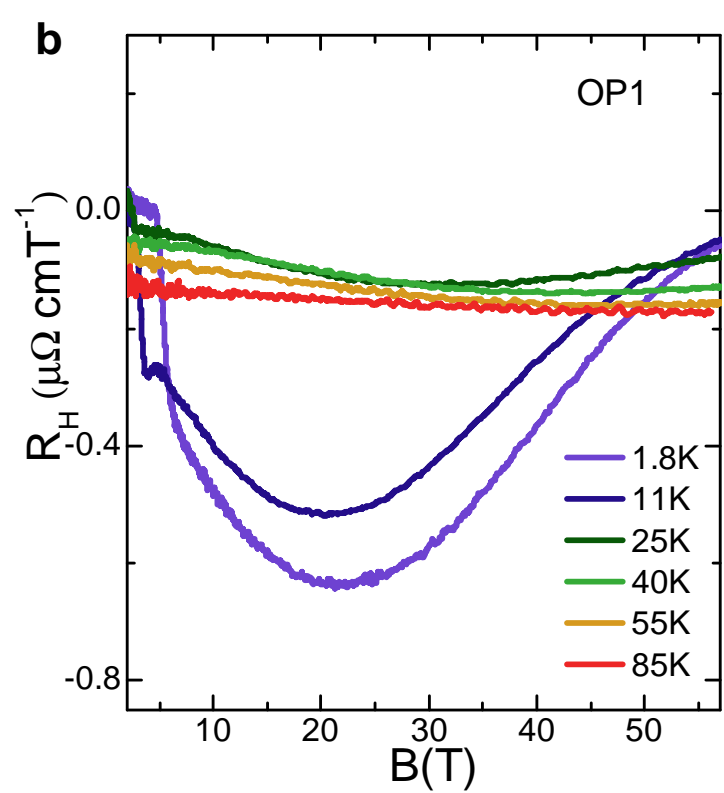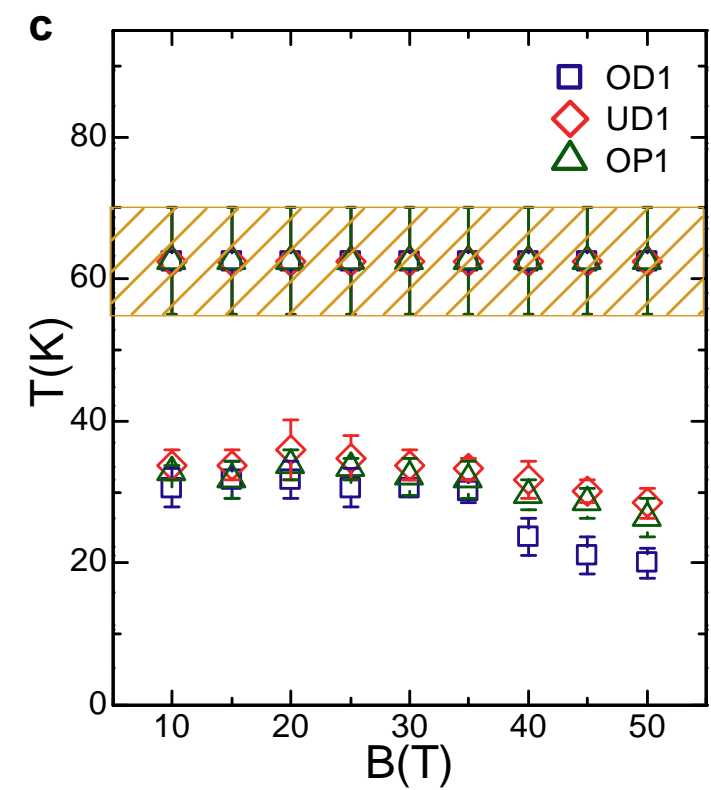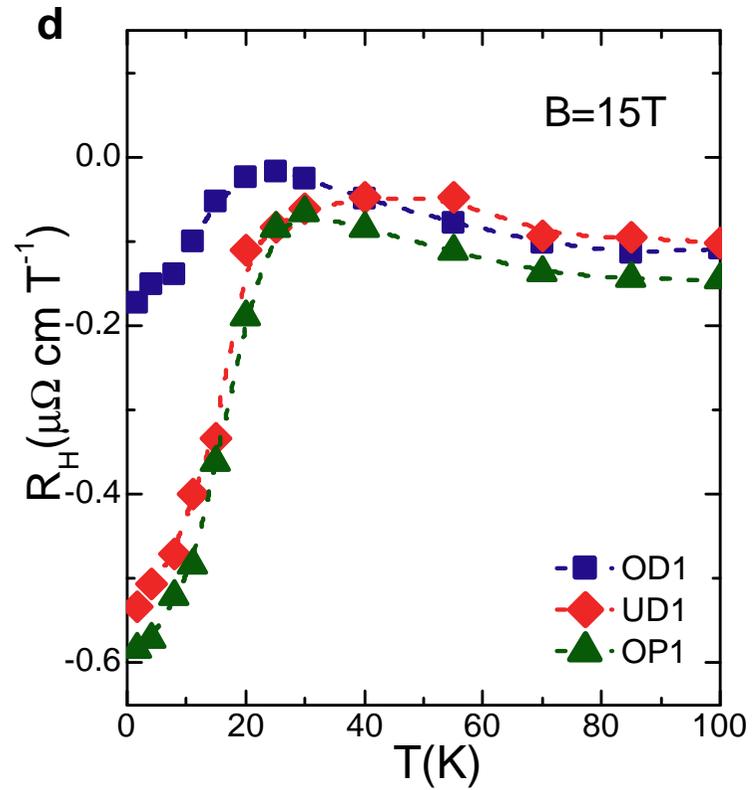

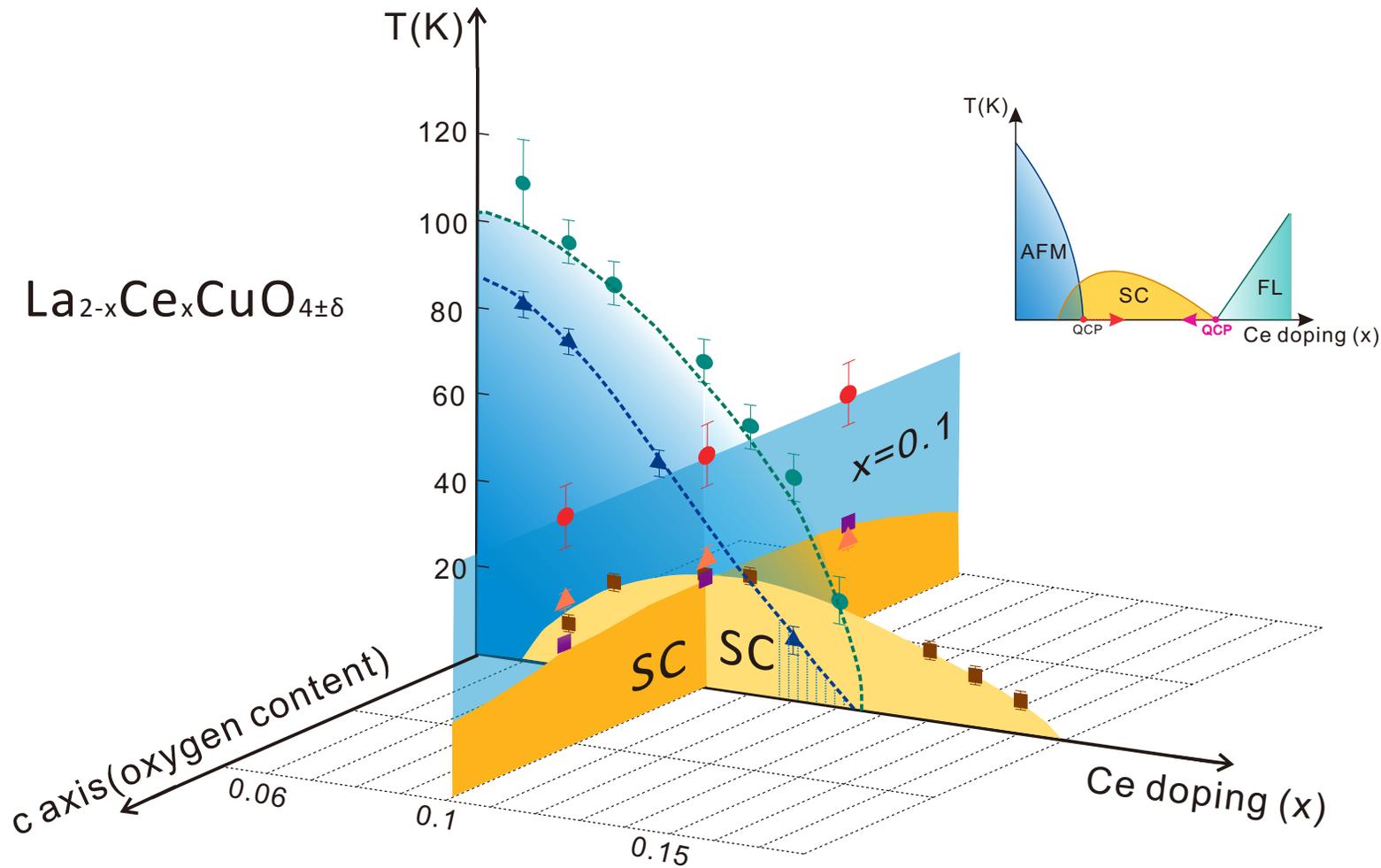

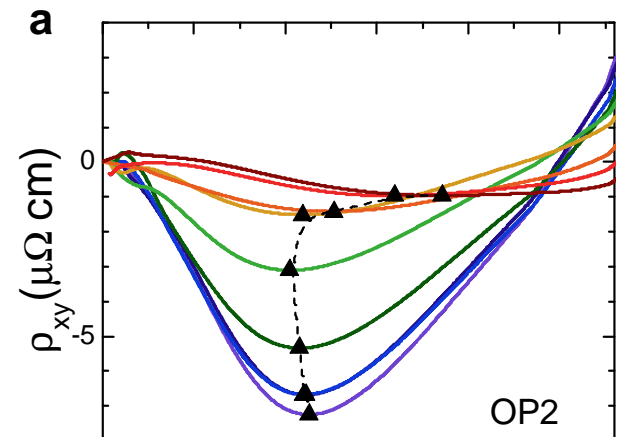
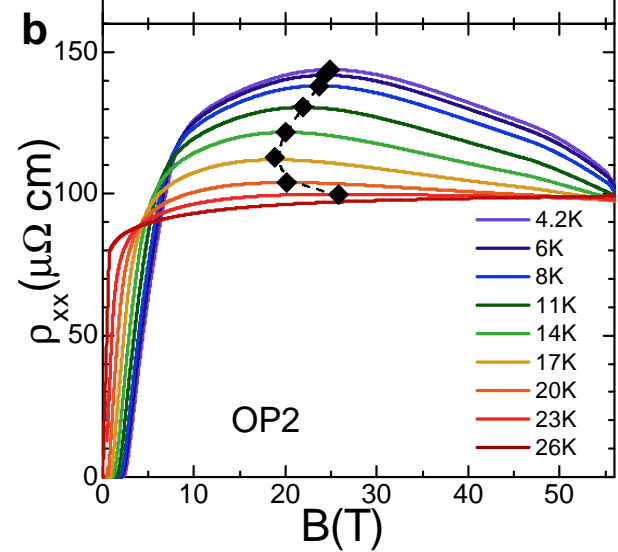
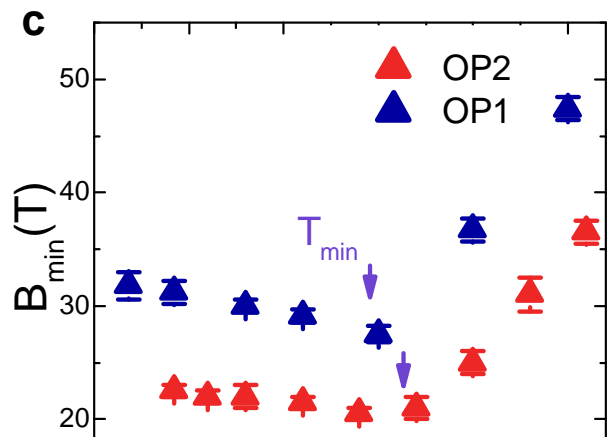
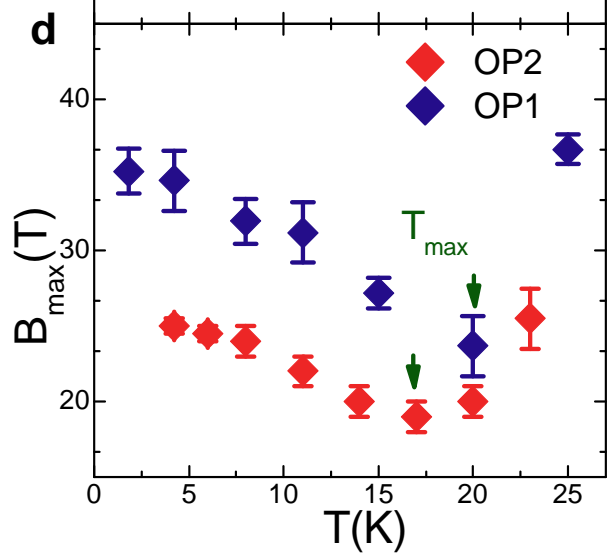
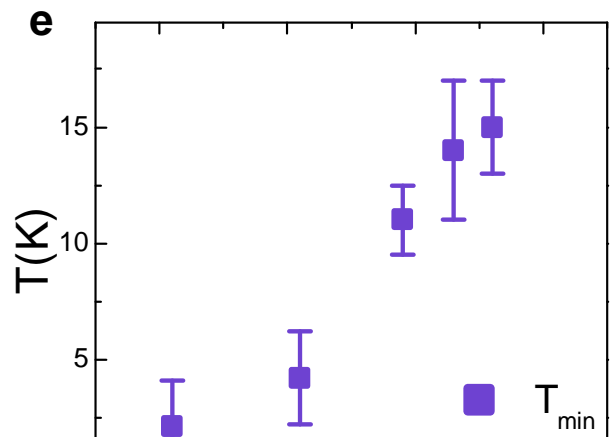
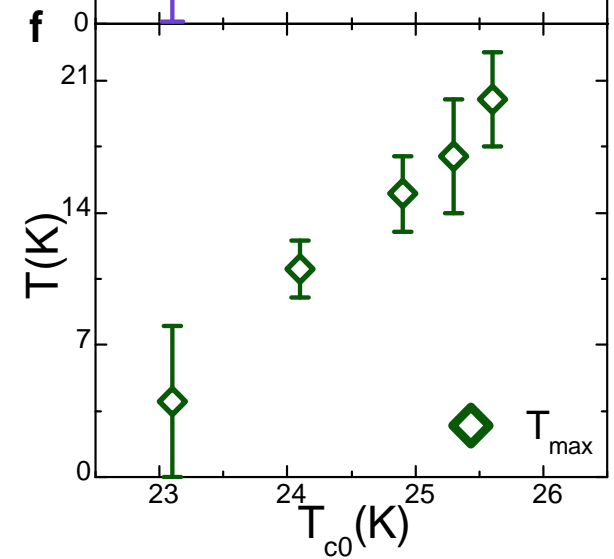

## Supplementary Information

**Sample syntheses and structural characterizations.** The c-axis-oriented $La_{2-x}Ce_xCuO_{4\pm\delta}$ ($x = 0.1$) thin films were deposited on the (00$l$)-oriented $SrTiO_3$ substrates by a pulsed laser deposition system. Since the $c$ and zero-resistance transition temperature $T_{c0}$ could be affected by annealing process, we used different annealing conditions to fabricate $La_{2-x}Ce_xCuO_{4\pm\delta}$ ($x = 0.1$) thin films with different $c$ and $T_{c0}$.

The structural measurements were finished by the instrument Smartlab (9000W) X-ray Diffractometer. Results of θ/2θ scan are shown in Figure S1. There exist only (00$l$)-oriented peaks of $SrTiO_3$ and $La_{2-x}Ce_xCuO_{4\pm\delta}$ ($x = 0.1$), which show that there is no impurity in our samples. We choose crystal plane (103) of $La_{2-x}Ce_xCuO_{4\pm\delta}$ ($x = 0.1$) to take the φ scan and achieve four nearly equal peaks after rotating sample along the normal line of crystal plane (103) from −200° to 160°. Then the reciprocal space mapping is taken for the crystal plane (103) of $SrTiO_3$ and crystal plane (109) of $La_{2-x}Ce_xCuO_{4\pm\delta}$ ($x = 0.1$). Results of φ scan and reciprocal space mapping reveal the high quality epitaxial growth of our samples. The structure parameters of the samples are calculated by the Bragg Equation ($2d_{00l}\sin\theta = \lambda$) from results of θ/2θ scan.

**Reciprocal space mapping (RSM).** The RSM scans were finished by the instrument Smartlab (9000W) X-Ray Diffractometer. The Schematic diagram of instrument is shown in Figure S6. There are four independent axes in the instrument, i.e., ω axis, 2θ axis, χ axis and φ axis. The ω axis and 2θ axis are along the $x$ axis, φ axis is along the $z$ axis, and χ axis is along to $y$ axis. The $yoz$ plane is the incident plane of x-ray.

We choose crystal plane (103) of $SrTiO_3$ and crystal plane (109) of $La_{2-x}Ce_xCuO_{4\pm\delta}$ to take the RSM scan. First, when 2θ is set as the diffraction angle of crystal plane (103) of $SrTiO_3$, the ω-scan is taken in order to find out the most appropriate value of ω (when the ω is set as this value, the strongest diffraction strength for crystal plane (103) of $SrTiO_3$ can be achieved). Second, ω/2θ scan is taken to confirm the scanning ranges of ω and ω/2θ. Here we assume that the ranges of ω and ω/2θ are $\omega_1 < \omega < \omega_2$ and $2\theta_1 < 2\theta < 2\theta_2$ and set the steps of ω scan and ω/2θ scan as δω and 2(δθ). Third, when the ω is set as $\omega_1$, the $\omega_1/2\theta$ is taken from $2\theta_1$ to $2\theta_2$. Then we set the ω as $\omega_1+\delta\omega$, and take $(\omega_1 + \delta\omega)/2\theta$ scan. The scan won't stop until the $\omega_2/2\theta$ scan is finished.

Results in the RSM scan are ω and ω/2θ, which can be transited into the expressions in the reciprocal space:
$$q_x = [\Delta\omega\cos\omega + (\Delta\theta − \Delta\omega)\cos(\varphi − \omega)]/\lambda$$
$$q_z = [\Delta\omega\sin\omega − (\Delta\theta − \Delta\omega)\sin(\varphi − \omega)]/\lambda$$
Here φ is the angle between the crystal plane (103) and the crystal plane (001) of $SrTiO_3$. $\Delta\omega = n\delta\omega$ and $\Delta\theta = m\delta2\theta$.

**Transport measurements in pulsed magnetic field.** Magnetotransport measurements in field up to 58 T were performed using a non-destructive pulse magnet with a pulsed duration of 60 msec at Wuhan National High Magnetic Field Center. Magnetoresistance and Hall resistance were measured simultaneously with a typical five probe method. Data for the up-sweeping and down-sweeping of the pulse field were in good agreement, thus we can exclude the heating effect of the sample by eddy current. Measurements with both positive and negative field polarities were made for all samples and measuring temperatures to completely eliminate the effects of contact asymmetries.

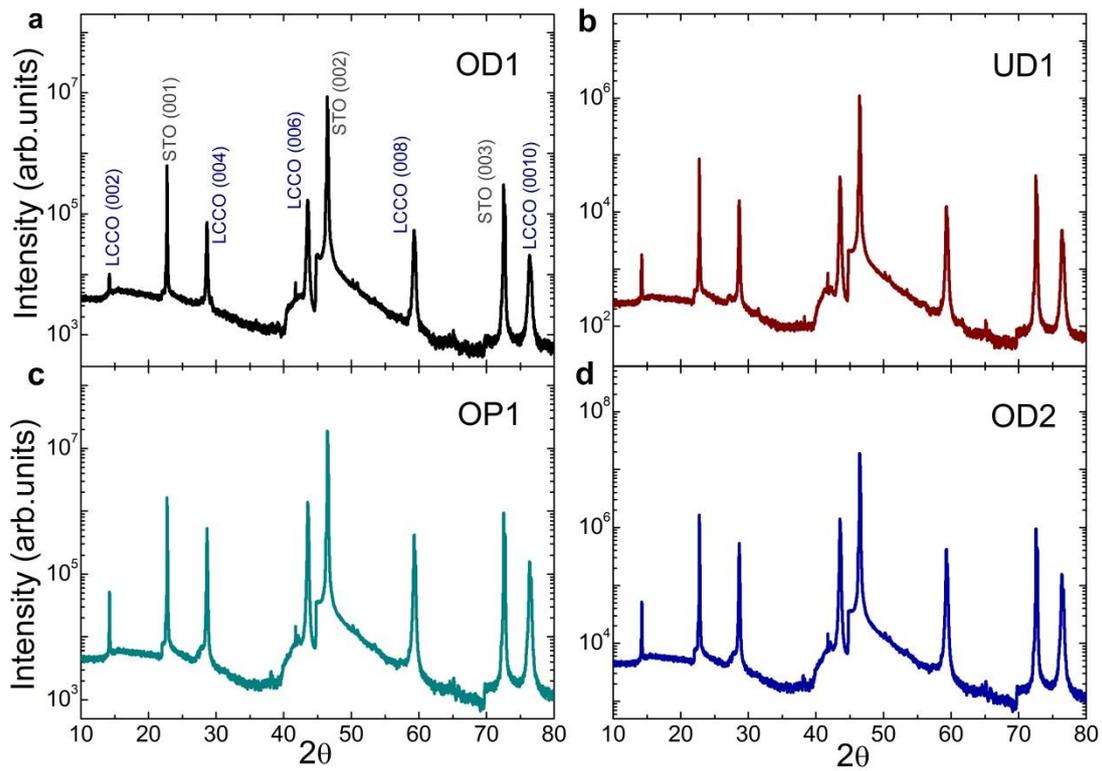

**Figure S1 | θ/2θ scan results of La$_{2-x}$Ce$_x$CuO$_{4\pm\delta}$ ($x$ = 0.1) thin films.** The θ/2θ scan results of samples **a,** OD1, **b,** UD1, **c,** OP1 and **d,** OD2. There are only (00$l$)-oriented peaks of SrTiO$_3$ and (00$l$)-oriented peaks of La$_{2-x}$Ce$_x$CuO$_{4\pm\delta}$($x$ = 0.1), which show that there is no impurity in our samples.

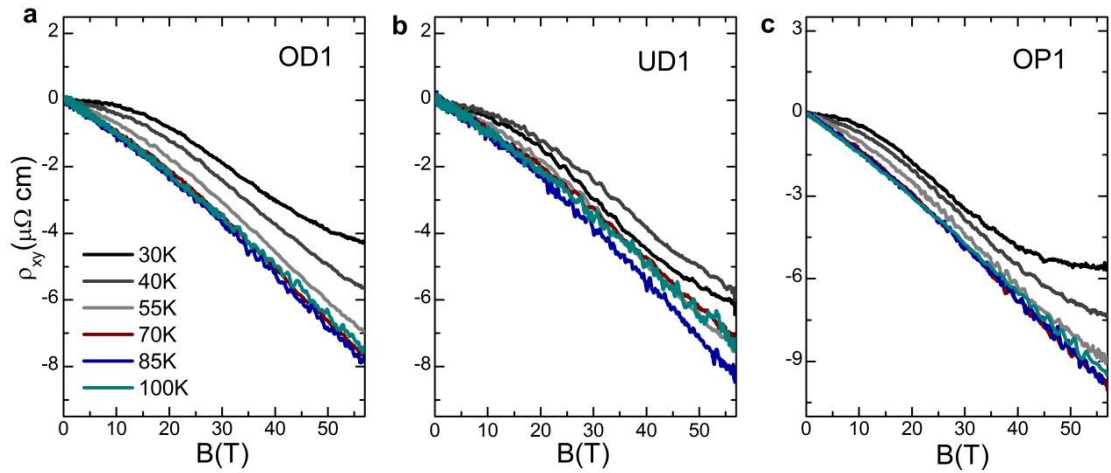

**Figure S2 | Field dependence of Hall resistivity $\rho_{xy}$ from 100 K to 30 K for samples, a,** OD1, **b,** UD1, and **c,** OP1. The relation between $\rho_{xy}$ and magnetic field is linearity at 100 K, 85 K and 70 K for all the three samples. But it turns to be nonlinearity when temperature is below 55 K. The characteristic temperature from linearity to nonlinearity is defined as $T_1$, which is nearly equal for the three samples.

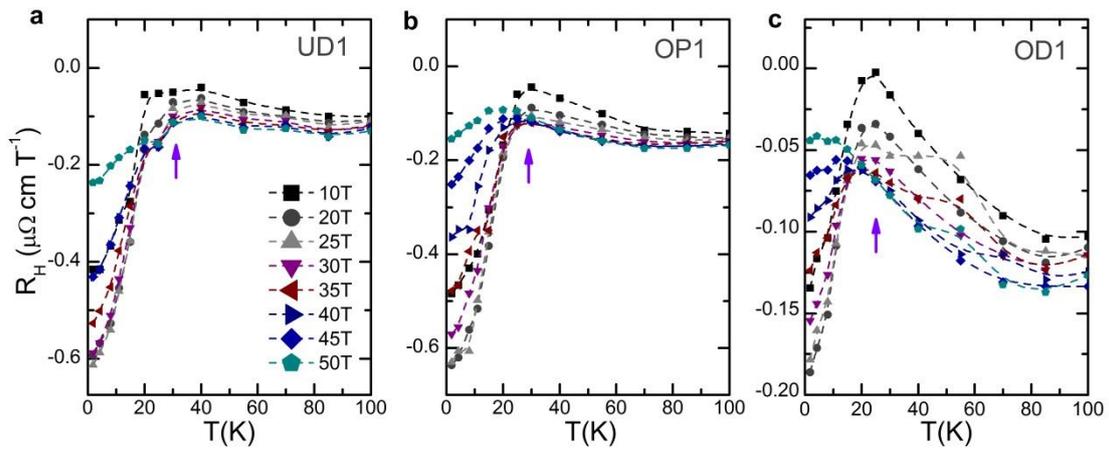

**Figure S3 | Temperature dependence of Hall coefficient $R_H$ from B = 10T to B = 50T for a,** UD1, **b,** OP1, **c,** OD1. There exists a kink at the $R_H(T)$ curves for the three samples. Temperature at the kink is nearly a constant when magnetic field is below 30 T. However, when magnetic field is above 30 T, it begins to change with increasing magnetic field.

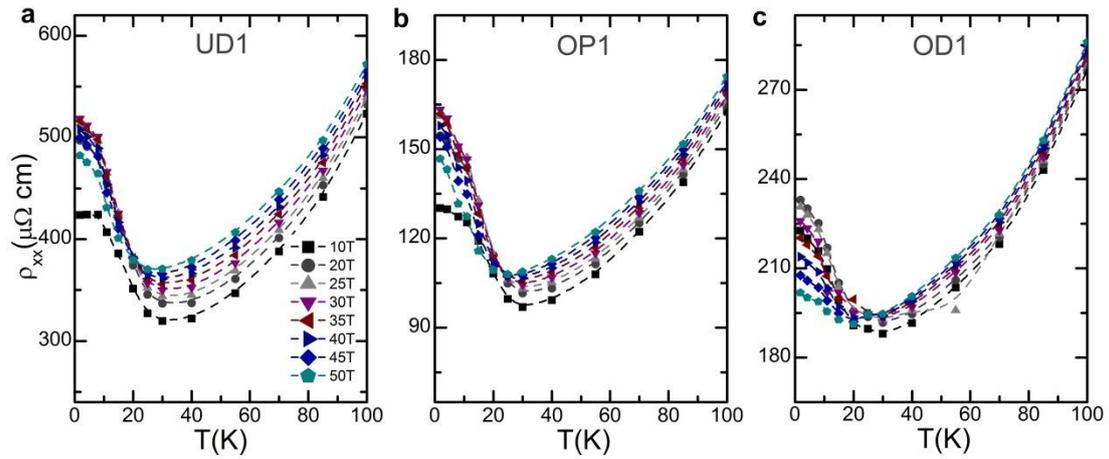

**Figure S4 | Temperature dependence of resistivity $\rho_{xx}(T)$ for a,** UD1, **b,** OP1 **and c,** OD1. The upturn toward insulatorlike behaviors exist in the $\rho_{xy}(T)$ curves from $B$=10 T to 50 T. When magnetic field is above 30T, the characteristic temperature where the $\rho_{xy}(T)$ shows a minimum begins to move with increasing magnetic field.

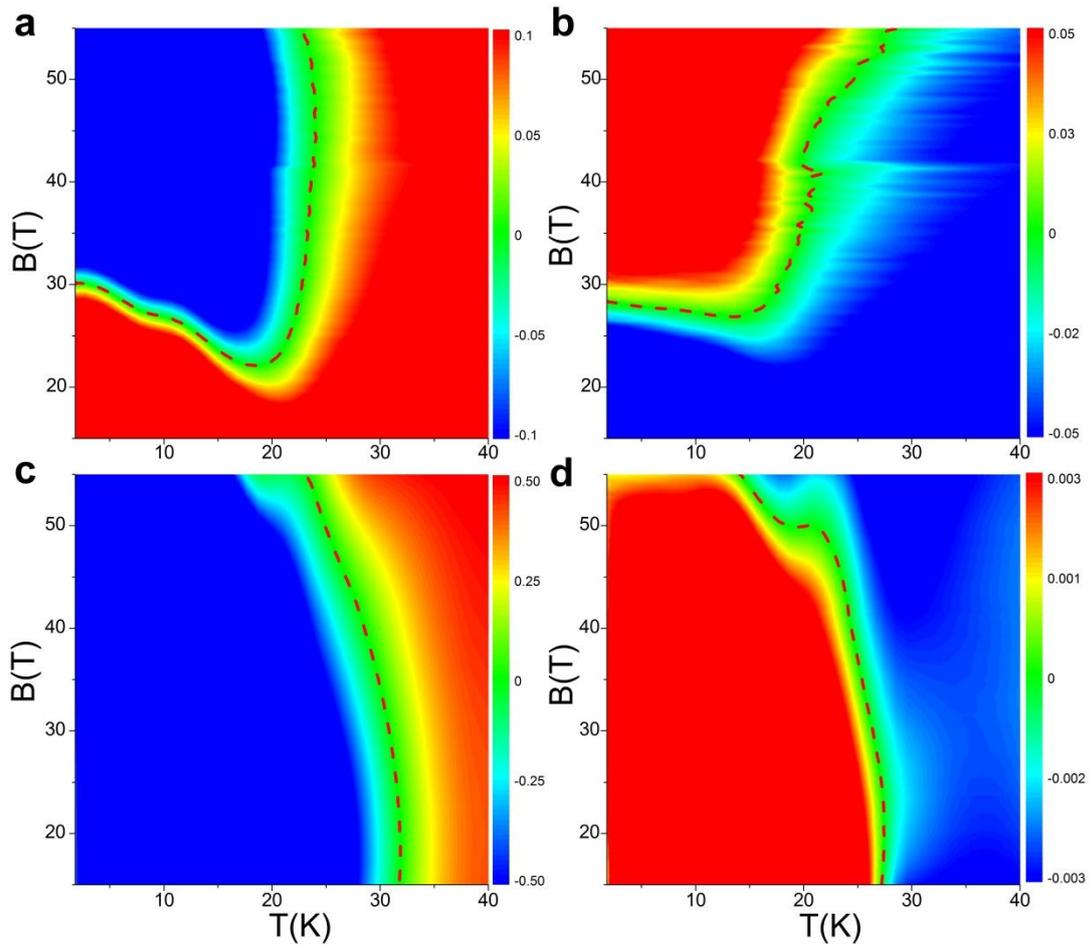

**Figure S5 | Contour maps of a,** d$\rho_{xx}(B,T)$ / d$B$. **b,** d$\rho_{xy}(B,T)$ / d$B$. **c,** d$\rho_{xx}(B,T)$ / d$T$. **d,** d$R_H(B,T)$ / d$T$ for sample OP2. The red dash line represents boundary ($B_{max}$) between negative MR regime and positive MR regime in **a**, the minimum ($B_{min}$) in $\rho_{xy}(B)$ curves in **b**, the connecting line of characteristic temperature where $\rho_{xx}(T)$ shows a minimum in **c**, and the schematic line of characteristic temperature at the kink of $R_H(T)$ curves in **d**. Both $B_{max}$ and $B_{min}$ first gradually decrease and then arise rapidly with increasing temperature. Characteristic temperatures where $\rho_{xx}(T)$ shows a minimum and at the kink of $R_H(T)$ curves both begin to change when magnetic field is above 30 T, while they are nearly a constant when B < 30 T.

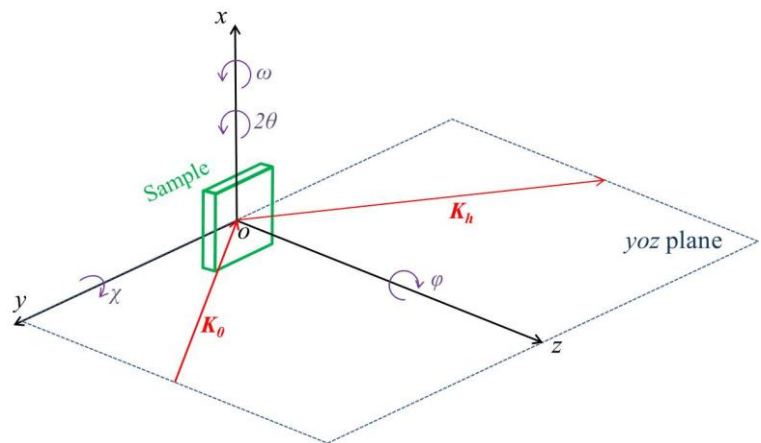

**Figure S6 | The schematic diagram of instrument used for reciprocal space mapping.** There are four independent rotation axes: $\varphi$-axis, $2\theta$-axis, $\omega$-axis and $\chi$-axis. $2\theta$-axis and $\omega$-axis are along the x-axis. $\chi$-axis is along y-axis, and $\varphi$-axis is along the z-axis. The incident plane of x-ray is the yoz plane.